\documentstyle[11pt,aaspp4]{article}
\begin{document}
\newcommand{\msun}{M_{\odot}}
\newcommand{\lsun}{L_{\odot}}
\newcommand{\zsun}{Z_{\odot}}
\newcommand{\kms}{\, {\rm Km\, s}^{-1}}
\newcommand{\cm}{\, {\rm cm}}
\newcommand{\erg}{\, {\rm erg}}
\newcommand{\mpc}{\, {\rm Mpc}}
\newcommand{\hz}{\, {\rm Hz}}
\newcommand{\seg}{\, {\rm s}}
\newcommand{\arcss}{\, {\rm arcsec}^{-2}}
\newcommand{\degs}{\, {\rm deg}^{-2}}
\newcommand{\pmag}{\, {\rm mag}^{-1}}
\newcommand{\nhi}{N_{HI}}
\newcommand{\lya}{Ly$\alpha$ }
\newcommand{\etal}{et al.\ }
\newcommand{\yr}{\, {\rm yr}}
\title{Searching for the Earliest Galaxies using the Gunn-Peterson Trough
and the Lyman Alpha Emission Line}
\author{
Jordi Miralda-Escud\'e$^1$ 
and
Martin J. Rees$^2$ }
\affil{$^1$ University of Pennsylvania, Dept. of Physics and Astronomy,
David Rittenhouse Lab.,
209 S. 33rd St., Philadelphia, PA 19104}
\affil{$^2$ Institute of Astronomy, University of Cambridge,
Cambridge CB3 0HA, UK}
\authoremail{ jordi@llull.physics.upenn.edu, mjr@ast.cam.ac.uk}

\begin{abstract}

  If the universe was reionized by O and B stars in an early population
of galaxies, the associated supernovae should have enriched the universe
to a mean metallicity $\bar Z = 10^{-5} (1+ n_{rec})$, where $n_{rec}$
is the mean number of times that each baryon recombined during the
reionization era. This is consistent with recent observations of the
metallicity in the \lya forest at $z\simeq 3$. The mean surface
brightness observable at present from the galaxies that produced these
heavy elements, in the rest-frame wavelengths $1216 {\rm \AA} < \lambda
\lesssim 2500 {\rm \AA}$, should be $\sim 10^{-6} (\bar Z/10^{-4})
(\Omega_b h^2/0.02) ~ {\rm photons}\, \cm^{-2}\seg^{-1} \arcss $. Most
of this radiation should be emitted at $z>5$, before reionization was
complete.

  These high-redshift galaxies may be detectable in near-infrared
photometric surveys, identifying them via the Gunn-Peterson trough
(analogous to the use of the Lyman limit cutoff to search for
galaxies at $z\sim 3$, where the \lya forest blanketing is smaller).
Their spectrum may also be characterized by a strong \lya emission
line. However, the spectra of galaxies seen behind intervening gas that
is still neutral should show the red damping wing of the Gunn-Peterson
trough, with a predictable profile that obstructs part of the \lya
emission.

  The low-mass galaxies formed before reionization might constitute a
distinctive population; we discuss the signature that this population
could have in the faint number counts. Although most of these galaxies
should have merged into larger ones, those that survived to the present
could be dwarf spheroidals.

\end{abstract}


\section{Introduction}

  Observations of large-scale structure have, over the last few decades,
strongly supported hierarchical models of gravitational instability
with ``cold dark matter''. Density fluctuations grow to non-linearity
first on small scales. High-density halos are formed, which then continue
to accrete low density matter around them and merge with each other to
form objects on progressively larger scales.
Evidence for this general picture includes the observed hierarchical
nature of galaxy clustering and the galaxy peculiar velocity field
(e.g., Strauss \& Willick 1995), the presence of substructure in galaxy
clusters showing that clusters are being assembled at the present time
from smaller units (e.g., Bird 1994),
the \lya forest at high redshift showing that structures on smaller
scales than the present clusters were collapsing in the past
(Rauch \etal 1997 and references therein), and the
CMB fluctuations (White, Scott, \& Silk 1994).
Even though the origin of the initial fluctuations
and the nature of the dark matter remain in the realm of speculation,
and the exact form of the linear power spectrum of density fluctuations
is uncertain, a generic consequence of this scenario is that the first
galaxies to form in the universe had very small scales and masses.
The earliest galaxies should have formed in the first collapsed objects
where the baryonic component was able to lose its energy through
radiative cooling, concentrating to the
center of dark matter halos and becoming self-gravitating, leading to
the formation of the first stars. In the absence of cooling, the gas
virialises and 
remains in hydrostatic equilibrium in dark matter halos and galaxies
cannot form (e.g., White \& Rees 1978 and references therein).

  As more massive structures form, the velocity dispersion of collapsed
halos increases, as does the temperature of the gas in these halos.
The gas that forms the first galaxies is, of course, initially neutral.
For $T \lesssim 5\times 10^3$ K, the only effective coolant in gas with the
primordial composition of hydrogen and helium is molecular hydrogen
(e.g., Tegmark \etal 1996). Molecular cooling almost certainly
played a role in forming the very first stars, but  the fraction of molecules
is always small and they should be rapidly photodissociated by UV
photons from the stars themselves (Haiman, Rees, \& Loeb 1997;
Ostriker \& Gnedin 1997). Radiative cooling is then suppressed until
temperatures $\sim 10^4$ K are reached, when atomic hydrogen provides
efficient cooling through line excitation and collisional ionization by
electrons in the Maxwellian tail of the energy distribution.
The cooling rate rises steeply with temperature, so the gas in this
regime contracts almost isothermally.

   The first-generation galaxies will eventually provide enough UV
photons to ionize all the diffuse gas in the universe (e.g., Couchman \&
Rees 1986); active nuclei could also start forming and contribute to the
process of ``reionization''(e.g., Arons \& Wingert 1972). The total
amount of star formation that precedes reionization obviously depends on
the IMF; it also depends on what fraction of the Lyman continuum from O
and B stars escapes into the intergalactic medium, rather than being
absorbed by gas in the star-forming regious themselves. After
reionization has occurred,  there is a universal UV background high
enough to maintain the HI fraction of any diffuse gas at a very low
level; this substantially reduces the cooling rate at temperatures below
$\sim 5\times 10^4$ K. This ionized diffuse gas can form the structures
responsible for the observed \lya forest (Cen \etal 1994,
Miralda-Escud\'e \etal 1996, hereafter MCOR, and references therein).
The higher entropy of the
ionized gas, together with the suppressed cooling rate, inhibit its
infall into halos with virial temperatures much below $10^5$ K (e.g.,
Efstathiou 1992; Thoul \& Weinberg 1996); dissipation is only efficient
in more massive objects, and presumably leads to the formation of most
galaxies that we have observed at all redshifts back to $z \lesssim 5$.

  The physics of the galaxy formation process is probably substantially
different depending on whether they form from neutral or from
photoionized gas, and this suggests that galaxies formed in two distinct
phases (e.g., Haiman \& Loeb 1997a). In this paper, we shall
designate ``Population A'' those galaxies formed from accretion of
photoionized gas, and ``Population B'' those galaxies formed in the
absence of an external ionizing background, so that photoionization
from external sources was negligible during the dissipation process.
Essentially, the Population B formed before the reionization epoch
and emitted the photons that
reionized the universe, and the Population A formed
afterwards, although in practice reionization is a gradual process
where different regions will be photoionized at different times, as
they are immersed in expanding cosmological HII regions that eventually
overlap.

  The distinction between these two populations is so far only a
theoretical concept, and our understanding of the effects of
reionization on galaxy formation is likely to remain limited until the
existence of the Population B (hereafter, Pop B) can be observationally
tested. In this paper we discuss the constraints that we can put on the
characteristics of the Pop B galaxies given our present knowledge of
the high redshift universe, and the prospects for their detection.
In a previous paper (Miralda-Escud\'e \& Rees 1997, hereafter MR97),
the possibility of detecting individual supernovae
from these galaxies was investigated, pointing out that although
supernovae at $z\gtrsim 5$ should be extremely faint,
their host galaxies could be even fainter,  given the
small masses expected for the Pop B galaxies. But as we shall see here,
the luminosities of Pop B galaxies may in fact be not much lower than
that of a supernova. Moreover, their detection is facilitated
because they should be exceedingly numerous,
and because of the distinct spectral signature
expected from starburst galaxies at high redshift:
the Gunn-Peterson trough and a possibly strong \lya emission line.

\section{The Brightness of Stellar Radiation Emitted at Reionization}

  It has recently been shown that the gas in the \lya forest was
already enriched to a heavy element abundance $Z \sim 10^{-2}\zsun$ by a
redshift $z=3$, from observations of absorption lines of CIV and other
species associated with \lya forest lines with $10^{14.7}\cm^{-2} < \nhi
\lesssim 10^{16} \cm^{-2}$ (Songaila \& Cowie 1996 and references
therein). Detailed calculations of the expected column densities of the
observed absorption lines, using hydrodynamic simulations of the \lya
forest and realistic
models for the spectrum of the ionizing background (Rauch, Haehnelt,
\& Steinmetz 1996; Hellsten \etal 1997) have shown that the carbon
abundance needed to reproduce the observations is $[{\rm C/H}]=-2.5$.
The metal abundances are similar to those of Population II
stars, where oxygen is the most abundant element and
is overabundant by a factor $\sim 2$ relative to carbon. With
$\zsun = 0.02$, the metallicity of the \lya forest is then
$Z\simeq 10^{-4}$.
As argued in MR97 this metallicity should be approximately the same
as the mean metallicity of the universe at $z=3$ if the \lya forest
contains most of the baryons (as is found to be the case in models of
structure formation for the \lya forest similar to those analyzed in
Hernquist \etal 1996 and MCOR), and if the \lya forest metallicity is
uniform. Actually, the metallicity probably increases with $N_{HI}$:
Hellsten \etal (1996) find that the absorbers with $\nhi < 10^{15}
\cm^{-2}$, which should contain a large fraction of the baryons, should
have lower metallicity than $[{\rm C/H}]=-2.5$, the value they need
to account for the CIV column densities of absorbers with higher
$N_{HI}$. The increase of metallicity with $\nhi$ is also predicted in
models of the enrichment of the IGM (Gnedin \& Ostriker 1997). This
can result in a lower mean metallicity. In fact, Songaila (1997) has
shown that by adding the column density of the CIV lines, the lower
limit to the mass of metals that is obtained is smaller by a factor
$\sim 10$ compared to the value we assume here. Thus, significant
uncertainties still remain on the mean metallicity. 

  The ratio of the mass of heavy elements ejected by a star to the
energy in ionizing photons emitted over the lifetime of the star, as
derived from models of stellar evolution and supernova explosions, turns
out to be about constant over the relevant mass range $10 \msun \lesssim
M \lesssim 50 \msun$ according to stellar evolution models (although the
possibility that some massive stars collapse to black holes without
ejecting heavy elements introduces some uncertainty), so given a mean
metallicity $\bar Z$ we can
predict the energy in ionizing photons that was emitted for each baryon
in the universe.  According to Madau \& Shull (1996), this energy is
$0.002 \bar Z m_p c^2$ per baryon. As discussed in MR97, for $\bar Z =
10^{-4}$ we find that $10$ ionizing photons were emitted per baryon by
the stars that produced the \lya forest heavy elements (assuming a mean
energy of 20 eV per ionizing photon). This is probably the radiation
produced by the Pop B galaxies, although it is certainly possible that
after the universe was reionized, some of the heavy elements produced in
the newly forming, more massive Pop A galaxies were also dispersed to
the intergalactic medium (hereafter, IGM) and contributed to the \lya
forest metal abundance.

  One of these 10 ionizing photons was used to reionize the universe.
The majority of the other photons must also have been absorbed
by neutral hydrogen. A fraction $f_i$ of the photons will be absorbed
internally, in the galaxies where the emitting stars were formed; the
rest should be absorbed after having escaped their original galaxies,
in other dense absorbing systems where protons can recombine many times
during the reionization epoch (similar to the observed Lyman limit
systems). Of the photons that were absorbed
internally, about 70\% resulted in the production of a \lya photon,
and the other 30\% resulted in emission in the two-photon continuum
from the 2s state (see Table 9.1 in Spitzer 1978; we assume that the
internal absorption takes place in
regions that are sufficiently optically thick to absorb essentially
all ionizing photons produced by direct recombination to the ground
state). The \lya photons will then be scattered many times in the
host galaxy, until they move to the wings of the \lya line and escape.
Some of them may be absorbed by dust before escaping, so we define
$f_d$ as the fraction of \lya photons that can eventually escape without
being absorbed by dust.
The mean comoving number density of these \lya photons is then
$7\, f_i f_d (\bar Z_{-4})n_b $, where $n_b$ is the comoving baryon
density and $\bar Z_{-4} = \bar Z/10^{-4}$. The mean surface
brightness of the \lya emission line from the Pop B galaxies is simply
obtained by multiplying the photon comoving number density by $c/4\pi$,
which yields $10^{-7}\, f_i\, f_d\, \bar Z_{-4} (\Omega_b h^2/0.02)\,
{\rm photons}\, \cm^{-2}\seg^{-1}\arcss)$. The fiducial value of
$\Omega_b h^2$ used here is close to that implied by a primordial
deuterium abundance $D/H \simeq 3\times 10^{-5}$, as reported by Tytler,
Fan, \& Burles (1996). 

  This determines the average surface brightness we receive in the sky
from these Pop B galaxies in the \lya emission line, or in other words,
the product of the number of Pop B galaxies per solid angle times their
individual, mean \lya flux. A much more uncertain question is the flux
from each individual galaxy; the uncertainty is essentially tied to the
efficiency of forming massive stars in collapsed dark matter halos as a
function of their velocity dispersion during the reionization epoch.
To see the range
of possibilities we might have for the main characteristics of Pop B
galaxies, we shall examine three possible scenarios:

(a) Star formation is highly efficient in all collapsed halos with
velocity dispersion $\sigma \gtrsim 15 \kms$, corresponding to the
virial temperature $T = \mu \sigma^2/k \simeq 10^4$ K where rapid
cooling by atomic hydrogen can start taking place. In this case,
reionization would take place as soon as the highest density peaks on
the scales of these low velocity dispersions collapse. 
As a specific example, let us take the top-hat spherical collapse
model for a halo, where a spherical region of mass $M$ turns around at a
radius $R_t$ when the age of the universe is $t_f/2$, and collapses at
time $t_f$ (e.g., Peebles 1980, \S 19), forming an isothermal halo
with $\sigma^2 = GM/R_t$ (obtained assuming that a mass $M$ is
contained within a radius $R_t/2$ in the final equilibrium
configuration). The halo mass is then $M=2^{3/2}/(2\pi G) \sigma^3 t_f
\sim 10^9 \msun (\sigma/20 \kms)^3\, (t_f/10^9 \yr)$.
If a fraction $f_b$ of the mass turns to stars
(with $f_b$ close to $\Omega_b$ for the high efficiency case) over a
timescale $t_s \sim 0.3 t_f$ of order the
free-fall time of the halo, the star formation rate is 
$0.3\msun\yr^{-1} (\sigma/20 \kms)^3\, (f_b/0.1)$. For a normal IMF, this
can yield a luminosity close to the peak luminosity of a supernova.
The implied metal production rate depends of course on the IMF, but is
$\sim 10^{-2}$ times the star formation rate when we calibrate it from
the observed yields in normal galaxies where the average metallicity is
$\sim$ 1\% after a large fraction of the gas has turned to stars.
Given the numbers
discussed above for the luminosity in \lya photons, which imply that
$10^5\, f_i\, f_d$ \lya photons are emitted for each baryon that is
expelled from stars in the form of heavy elements, the \lya luminosity
from this efficient, rapid starburst in a Pop B galaxy would be
$10^{52}\, f_i\, f_d\, (\sigma/20\kms)^3\, (f_b/0.1)\, {\rm photons}\,
\seg^{-1}$.
The corresponding \lya flux at present (for $\Omega=1$ and
$H_0=50 \kms\mpc^{-1}$) is
$5\times 10^{-7}\, f_i\, f_d\, (\sigma/20\kms)^3\, (f_b/0.1)\,
\left[1-(1+z_f)^{-1/2}\right]^{-2}\, (1+z_f)^{-1} \,
{\rm photons}\, \cm^{-2}\seg^{-1}$. If the sources that reionized the
universe have these characteristics, then we can use the mean \lya
surface brightness we obtained earlier to conclude that the number
density of these sources in the sky should be $\sim 1$ per arcsec$^2$
for the fiducial numbers we have chosen, and a formation redshift
$1+z_f\sim 10$. The redshift $z_f$ should of course depend on the
detailed model for the amplitude of the primordial fluctuations on
small scales.

  The main problem with this scenario is that, as pointed out by several
authors (e.g., Dekel \& Silk 1986), the energy from supernova
explosions in galaxies of such low velocity dispersions should easily
be able to expel the gas from the halo when only a small fraction of
the gas has turned to stars. In fact, the enriched gas needs to be
expelled at some stage to spread the heavy elements through the IGM.
Although one could imagine the supernovae
taking place in a very dense medium where the energy was quickly
radiated as soft X-rays, it seems likely that the supernova explosions
will reduce the efficiency of star formation by a large factor, due to
the heating and expulsion of gas. Indeed, it has been argued that this
reduction of the star formation rate is necessary to prevent too many
baryons from turning to stars in small galaxies at early times
(e.g., Navarro \& Steinmetz 1997). This leaves us with two other types
of Pop B galaxies that could be the sources of the reionization photons,
our scenarios (b) and (c).

(b) Reionization is caused mostly by low-mass objects similar to case
(a), but only a small fraction of the gas in each object turns to
stars before the gas is expelled in a wind. Thus, the parameter
$f_b \ll \Omega_b$, so each Pop B galaxy emits fewer photons and a
larger number of them have to form, which therefore have to originate
from lower amplitude peaks. The luminosity of each galaxy would be
reduced proportionally to $f_b$ due to the smaller mass of stars.
This might be partly compensated by having starbursts of very short
duration (in principle, starbursts could occur over a time much shorter
than the free-fall time through the dark matter halo if only a small
fraction of the halo gas turns to stars).

(c) The efficiency of massive star formation in the first halos where
atomic cooling takes place is so low that not enough photons are emitted
to complete the reionization until more massive galaxies (with $\sigma
\gtrsim 100 \kms$ collapse and form stars more efficiently. In this
case, the Pop B galaxies responsible for the reionization would have
relatively large masses, well above the Jeans mass for
photoionized gas, so reionization might be of little consequence for the
process of galaxy formation, and the Pop B galaxies might not have any
observational characteristic that would distinguish them from Pop A
galaxies (this is because photoionization has little effect on the
cooling rate in halos with velocity dispersion $\sigma \gtrsim
100 \kms$; see Thoul \& Weinberg 1996, Navarro \& Steinmetz 1997).

  The inefficiency of star formation in low $\sigma$ galaxies might
arise from two different reasons. One is the rapid expulsion of the gas
after a very small number of stars have formed, as in case (b). The
other possibility is that the gas is not expelled at all, but instead
forms rotationally supported disks by radiative cooling and dissipation,
and then the gas in these disks stays in atomic form and does not
effectively turn to stars. This could happen because the low-metallicity
gas might not cool to the cold phase of the interstellar medium (see
Corbelli \& Salpeter 1995). Staying at a temperature $T\simeq 5000$ K in
a disk with a low circular rotation velocity, the disks might be
gravitationally stable according to Toomre's criterion, preventing the
formation of molecular clouds. These stable disks might persist until
their host halos merge within larger systems, when the disks would be
disrupted and the gas would again dissipate and form a new, more massive
disk, with a higher efficiency of star formation.
A low star formation efficiency in low $\sigma$ galaxies could also be
part of the solution to the common problem that hierarchical models
have in producing too many low-mass galaxies (e.g., Kauffmann,
Guiderdoni, \& White 1994).

  We need to point out here that moderately massive galaxies are
likely to form during reionization, even in cases (a) and (b), because
the power spectrum at small scales in cold dark matter models
generically flattens to a slope close to $n=-3$,
so the amplitude of fluctuations decreases very slowly with scale.
As an example, let us consider the standard cold dark matter
model with $h=0.5$, and the rms density fluctuation normalized at
present to $\sigma_8=0.7$ on spheres of $8 h^{-1} \mpc$. This model has
been shown to be approximately in agreement
with the observed characteristics of the \lya forest
(Hernquist \etal 1996, Dav\'e \etal 1997, Rauch \etal 1997),
and therefore should probably have about
the right amplitude of density fluctuations on small scales.
The three solid lines in Figure 1 show the velocity dispersion
(or virial temperature $T_{vir} = \mu \sigma^2/k$) of halos having
collapsed from (1,2,3)-$\sigma$ fluctuations as a function of
redshift (assuming $\delta_c=1.69$ for the
critical value of the linear overdensity that corresponds to the
formation of a galaxy, as in the spherical collapse model). The
dashed lines indicate constant halo mass.
If, for example, reionization was complete at $z\simeq 9$, when halos
with $\sigma=20 \kms$ would just be collapsing from $2-\sigma$ peaks,
at the same epoch we would have halos with $\sigma=75 \kms$ (with a
mass $\sim 50$ times larger than halos with $\sigma = 20 \kms$)
collapsing from $3-\sigma$ peaks on this larger scale. In a Gaussian
theory, the $3-\sigma$ peaks should contain $\sim 10\%$ as much mass
as the $2-\sigma$ peaks at a fixed epoch, so we see that the mass
distribution of the Pop B galaxies should probably extend well above
the minimum mass for efficient atomic cooling. Notice that the very
large range in mass of collapsed objects at high redshift is a generic
consequence of the flattening of the power spectrum on small scales,
due to the fact that the universe was radiation dominated when these
small scales entered the horizon. Objects from 1-$\sigma$ peaks,
which contain most of the mass, probably never cooled before
reionization because they were ionized by the radiation from more
massive objects, and could only start collapsing at $z\simeq 3$,
at $\sigma \simeq 30 \kms$, when the cooling rate was again fast enough.
But objects from 3-$\sigma$ peaks at $z=9$ could already have total
masses over $10^{10}\msun$.

\section{Detecting Galaxies at $z> 5$}

  We now discuss the prospects for detection of this population of
galaxies at high redshift.
First, we notice that given the mean surface brightness of
\lya photons derived previously, the surface brightness in the UV
continuum also follows with relative little additional uncertainty,
because the stars emitting most of the light in the range $1216 
{\rm \AA} < \lambda \lesssim 2500$ \AA are also mostly very young
stars. Given the typical UV spectrum of a starburst (e.g., Fig. 1 in
Bruzual 1983), we infer that for every \lya photon produced from
ionizing photons there should be $\sim 10$ UV continuum photons
(because the energy emitted in ionizing photons is $\sim 1/3$ of the
UV energy emitted to the red of \lya in a young starburst, and $1/3$
of the energy in ionizing photons is converted to \lya photons). This
implies that the mean surface brightness from all Pop B galaxies in
the rest-frame UV continuum is $10^{-6}\, {\rm photons}\,\cm^{-2}\seg^{-1}
\arcss$. The value of this mean surface brightness agrees with the
calculation by Haiman \& Loeb (1997b), who used also the metallicity
of the \lya forest as a constraint and made a more detailed model of
the stellar population and the distribution of galaxy luminosities.

  The main spectral feature that should identify
any such galaxies at $z \gtrsim 5$ is a sharp break of the UV continuum
at the \lya wavelength, due to the Gunn-Peterson trough
(Gunn \& Peterson 1965);
in addition, the \lya emission line may be present, depending on
internal dust absorption and scattering of the \lya photons
(see the next section).
Notice that the redshift at which the IGM was reionized
is not highly relevant regarding the presence of the Gunn-Peterson
trough, because even if the medium was reionized at $z \gg 5$, we know
that the flux decrement caused by the \lya forest reaches a factor of 2
at $z\simeq 4$ and grows rapidly with redshift. Thus, the technique of
identifying galaxies at $z\simeq 3$ from the Lyman continuum break
(Guhathakurta \etal 1990; Steidel \etal 1996) should be replaced by
the Gunn-Peterson trough at $z \gtrsim 5$ (see Madau 1995 and Madau
\etal 1996 for a careful analysis of the effects of the \lya forest
on galaxy colors), although some residual flux between the \lya line
and the Lyman limit would still be present up to $z\simeq 7$ if the
reionization occurred earlier, and the Lyman limit discontinuity may
then be used in conjunction (see Loeb \& Haiman 1997).

  The mean surface brightness in the rest-frame UV continuum from the
Pop B galaxies can also be expressed as $S_\nu = 6\times 10^{-33}
\erg\cm^{-2}\seg^{-1}\hz^{-1}\arcss = 32\, {\rm AB}\arcss$ (where AB
denotes AB magnitudes in the band where the UV continuum is observed;
notice that in these units, the derived surface brightness is also
independent of redshift because the $(1+z)$ factor due to the redshift
of the photons is cancelled since we express the surface brightness in
energy per unit frequency). Here we have adopted the values
$\Omega_b h^2 = 0.02$, $\bar Z = 10^{-4}$; the surface brightness is
of course proportional to these two quantities.
If we now take the example of case (a) in the last
section, where there is one Pop B galaxy per arcsec$^2$ at redshifts
$z_f \sim 10$ with $\sigma\simeq 20 \kms$ (corresponding to 2-$\sigma$
peaks in Fig. 1), each such galaxy would have an AB magnitude of 32.
However, as mentioned earlier, realistically these high-z galaxies
should have a wide range of luminosities, corresponding to the range of
their masses. The dotted lines in Figure 1 indicate the AB magnitudes
of galaxies in halos during a starburst, under the same assumption as
in \S 2 that all the baryons turn to stars over the time $t_s \sim
0.3 t_f$. The AB magnitude is for rest-frame wavelengths in the UV
continuum, longward of \lya . Thus, galaxies forming from $3-\sigma$
peaks could have AB magnitudes of 28 even at $z\sim 10$.
Even though these would be $\sim 500$ times more rare than $2-\sigma$
peaks at the same redshift, the number density could still be close to
one galaxy per square arc minute. Our scenarios (b) or (c)
would be more optimistic because, for a fixed epoch of reionization, a
model with a higher amplitude of density fluctuations
and a greater number of massive halos at high redshift would be
required due to the lower overall efficiencies of star formation and
production of ionizing photons.

  These high-redshift galaxies will probably be detectable soon.
Detection with ground-based telescopes is hindered by the high sky
background in the infrared, but may still work at $z \lesssim 6$, and
in \lya searches where special wavelengths with low atmospheric
emission are chosen.
The Keck telescope can detect point sources to AB magnitudes $R < 28$,
and $I < 27$, in a night of observing time
(M. Rauch 1997, priv. communication; Cohen 1995a,b).
The magnitude limit could be improved with adaptive optics.
With space instruments,
the faintest galaxies in the HDF reach to $I \simeq 28.5$ (Williams
\etal 1996). The {\it New Generation Space Telescope} could image
galaxies to AB magnitudes $\sim 31$ in the near-infrared, and should
be able to detect galaxies to much higher redshift
(Mather \& Stockman 1996),
with a much higher number density than has been seen so far.
The prospect for detecting the high-redshift galaxies responsible for
the enrichment of the \lya forest has also been analyzed in Haiman
\& Loeb (1997b), Loeb \& Haiman (1997), Loeb (1997) and Cen (1997).

  In general, the lensing magnification in rich lensing clusters may be
used here to stretch the magnitude limit (see also Cen 1997). As an
example, a lensing cluster with an Einstein ring radius $b=30''$ should
magnify to $A> 10$ an area of $\sim 30$ arcsec$^2$ in the source plane.
In the example used above, about 30 galaxies with $AB=32$ could be in
this area, which would be magnified to $AB=29.5$. Magnified images of
high-redshift galaxies should characteristically appear in pairs around
the critical lines, in a region that can be predicted from lensing
models (see Miralda-Escud\'e \& Fort 1993; Kneib \etal 1996, and
references therein), so this should help in their identification. These
numbers indicate that a new deep field (similar to the HDF) imaged with
HST in a rich cluster, adding also the H and J filters in the
near-infrared, might well identify several galaxies at $z > 5$. In fact,
the largest redshift object known at present (at $z=4.92$) is already a
gravitationally lensed galaxy (Franx \etal 1997).

  We have assumed in this discussion that these high-redshift galaxies
would be sufficiently small to remain unresolved. Resolved object would
need to have higher fluxes to be detected, since the detection is
limited by the sky background. The scale-lengths of the Pop B galaxies
are expected to be very small. The radius of a collapsed halo scales as
$\sigma\, t_f$. Compared to present day galaxies, the Pop B galaxies
at $z> 5$ should have a formation time smaller by at least a factor
$\sim 10$, and a velocity dispersion that is also smaller by a factor
$\sim 5$. We can reasonably assume that the ratio of the size of the
region where stars form to the radius of the collapsed dark matter halo
is similar for all types of galaxies, if the processes of dissipation
and distribution of the angular momentum that determine the size of
rotating disks (Fall \& Efstathiou 1980), or the onset of star formation
in spheroidal components when a gas cloud becomes self-gravitating and
fragmentation occurs, are similar in galaxies of different masses.
Therefore, given the typical scales of a few Kpc for present-day
galaxies, the scales of Pop B galaxies are likely to be smaller than
100 pc, corresponding to angular sizes $\sim 0.01$ arcsec. This would
remain unresolved even with NGST.

\section{The Lyman Alpha Emission Line}

  Another possible way to detect the faint Pop B galaxies is to search
directly for the \lya emission line. As we discussed before,
for a normal starburst spectrum we expect $\sim$ 10\% of the UV photons
to be in the \lya emission line if dust absorption is not important.
Therefore, if the sensitivity for detecting galaxies is still
limited by the sky background for emission-line searches, 
the width of the line should be $\Delta \lambda/\lambda < 0.01$ to
allow detection on a shorter time than for the UV continuum. The
width of the \lya line in emission from a region of neutral gas with
column density $\nhi = 10^{22} N_{22} \cm^{-2}$, and velocity dispersion
$\sigma = 10 \sigma_6 \kms$ is $\Delta \lambda / \lambda \simeq
2\times 10^{-3} (N_{22}\sigma_6)^{1/3}$ (Harrington 1973; this formula
is valid for a very optically thick system where photons are scattered
many times in the damping wing before escaping from the system). Thus,
the sensitivity of emission-line searches might be at best similar to
searches for the UV continuum, if the sky background were smooth in
wavelength.

  From the ground, line-emission searches have the advantage that a
wavelength of particularly low atmospheric emission can be selected,
so the line-emission search can then be substantially more sensitive
than a photometric one. The deepest \lya emission surveys so far have
reached limits of $3\times 10^{-6} {\rm photons}\cm^{-2}\seg^{-1}$ at
$z \sim 4$ (Thompson, Djorgovski, \& Trauger 1995), and $10^{-4}
{\rm photons}\cm^{-2} \seg^{-1}$ at $z\sim 7$ (Parkes, Collins, \&
Joseph 1994). With our assumed ratio of 10 UV continuum photons for
each \lya photon (and ignoring the possibility of dust obscuration),
these limits correspond to AB magnitudes of 28 and 24.5 in Figure 1,
respectively. At $z\sim 4$, the population of luminous starbursts
must be obscured by dust, as shown by the recent discovery by
Steidel \etal (1996) of the Lyman limit dropout galaxies with
$AB \sim 26$ at this redshift. But the lower mass galaxies at
higher redshift might have stronger \lya emission lines. Some
examples of high redshift galaxies with strong \lya emission have
been found (Hu \& McMahon 1996; Franx \etal 1997).

  The \lya emission line may be suppressed due to internal dust
extinction. In addition,
any photon emitted in the blue side of the \lya line will be scattered
in the surrounding IGM, reducing the strength of the line by a factor
of 2. For a galaxy observed when the surrounding IGM is still mostly
neutral, even photons on the red side can be scattered due to the
damping wing of the Gunn-Peterson trough. Assuming a neutral IGM with
uniform density, the Gunn-Peterson optical depth is
$\tau_{GP} = 2.1\times 10^5\, [\Omega_b h (1-Y)/0.03]\, [(1+z)/6]^{3/2}$
(where $Y$ is the primordial helium abundance).
The optical depth at a wavelength $\Delta \lambda$ to the red of the
\lya line is

\begin{equation}
 \tau(\Delta\lambda) = {\tau_{GP}\, R_\alpha\over \pi } \,
\int_{\Delta\lambda/ \lambda}^{\infty}\,
{dx\over x^2 + R_{\alpha}^2 } = \tau_{GP}/\pi\, 
{\rm arctan} \left[ R_{\alpha} (\Delta\lambda/\lambda)^{-1} \right] ~ ,
\end{equation}
where $R_{\alpha} = \Lambda/(4\pi\nu_{\alpha})=2.02\times 10^{-8}$,
$\Lambda=6.25\times 10^8 \seg^{-1}$ is the decay constant for the \lya
resonance, and $\nu_{\alpha} = 2.5\times 10^{15} \hz$ is the frequency
of the \lya line. This expression can be approximated to
\begin{equation}
\tau (\Delta\lambda) =
\tau_{GP} R_{\alpha}/\pi (\Delta\lambda/\lambda)^{-1} =
1.3\times 10^{-3}\, [\Omega_b h (1-Y)/0.03] [(1+z)/6]^{3/2}
(\Delta\lambda / \lambda)^{-1} ~ .
\label{dampp}
\end{equation}
This width of the damped absorption from the IGM is of similar breadth
as the emission lines we may expect from starburst galaxies, so the
red side of the \lya emission line should be partially suppressed by
the IGM in a galaxy observed before the reionization.
It is possible that the ionizing radiation of the galaxy itself would
have ionized the surrounding IGM, but this is unlikely in a source
with a strong \lya emission line because most of the ionizing photons
need to be absorbed to produce the line.
The edge of the Gunn-Peterson trough should have the shape given by
equation (\ref{dampp}), except for the fact that the inhomogeneity of
the IGM may alter the profile (in particular, the presence of a halo of
gas accreting on the galaxy may substantially increase the column
density contributing to the damped profile from gas at a redshift close
to that of the galaxy). This damping wing is studied in more detail
in Miralda-Escud\'e (1997).


\section{Effect of the Population B on the Faint Galaxy Number Counts}

  We have calculated in this paper that the galaxies where the heavy
elements observed in the \lya forest at $z\simeq 3$ were synthesized
have to contribute a mean surface brightness of $32 \, {\rm AB}\arcss$
to the galaxy number counts. The majority of these galaxies should be at
redshifts higher than any galaxy redshifts determined so far
($z \gtrsim 5$), due to two reasons: first, it is
unrealistic that the massive stars could have formed in a highly
synchronous fashion, shortly before the redshift where the heavy
elements are observed; and second, the dispersion of the heavy elements
through the \lya forest should take a substantial fraction of a Hubble
time, because the enriched gas is not likely to be expelled from the halos
around galaxies at a velocity much higher than the escape speed.

  Let us now compare this predicted mean surface brightness for the
Population B with the mean surface brightness of the faintest galaxies
observed so far. We shall do this in the I and K bands. In the I band,
the faintest galaxies in the HDF are at $I\simeq 28.5$, reaching a
number density at this magnitude of $5\times 10^5 \degs\pmag$ (Williams
\etal 1996). The slope of the counts at this magnitude is $\sim 0.2$
(meaning that the counts are increasing as $10^{0.2 I}$). If we assume
that this slope remains constant at all fainter magnitudes, the total
mean surface brightness from galaxies with $I>28.5$ is $31.2 \,
{\rm AB}\arcss$. Because this is only a factor $2$ brighter than the
expected value for the Pop B galaxies, we can conclude that if all the
Pop B galaxies were at $z<6$ (and therefore still observable in the
I band), then either these Pop B galaxies should be the objects
detected just at the magnitude limit in the HDF, or the slope of the
I counts should rise again at fainter magnitudes to a value greater than
$0.4$, to yield a mean surface brightness comparable to the one
contributed by the galaxies with $I\sim 29$.
The first possibility is probably ruled out from the
work of Madau \etal (1996), who find that only a small fraction of the
faint galaxies in the HDF can be at $z > 3.5$ (see their Fig. 8a).

  Actually, the I counts do not need to steepen again at $I > 29$,
as long as most of the surface brightness inferred for the Population B
comes from galaxies with $z> 6$, since the Gunn-Peterson trough will
then cause these galaxies to drop out of the I band. But the same
argument can be applied at longer wavelengths as better observations
become available in the J, H and K bands in the future, from HST and
NGST. If the counts were observed to flatten to a shallow slope, and
the mean surface brightness from the faintest observed galaxies were
already less than $32 \, {\rm AB}\arcss$, while the redshifts of these
galaxies were still not significantly higher than the redshift at which
the heavy elements were present in the \lya forest, then we could infer
that the slope of the galaxy counts should steepen again; in other words,
there should be a ``second hump'', a maximum in the contribution to the
mean surface brighness as a function of galaxy magnitude, due to the
Pop B galaxies. The presence of a ``second hump'' would be evidence that
the Populations A and B are really distinct, as in the cases (a) and (b)
discussed in Section 2 (unless dust obscuration could cause the dip
between the two ``humps'' in the galaxy counts).

  The present observations in the K band from the ground are not yet
deep enough to put significant constraints along this line of argument.
The faintest counts determined so far yield a density $2.2\times 10^5
\degs\pmag$ at $K=26.4$ (Djorgovski \etal 1995; the AB magnitude
$K=26.4$ corresponds to $K=23.5$ in the Johnson system). Assuming
a constant slope of $0.3$ for the faint counts (the observed slope in
the range $23 < K < 26.4$), the
mean surface brightness due to galaxies with $K>26.4$ should be
$29.2 \, {\rm AB}\arcss$, still larger by a factor $\sim 10$ compared
to the expected value for the Population B. Thus, the Pop B galaxies
could appear in the K counts at $K \gg 26.4$ without requiring a
steepening of the slope. 

\section{Dwarf Spheroidal Galaxies: Remnants of the Population B?}

  If the formation of galaxies occurred indeed in two distinct
populations, due to the reionization of the IGM, we might then expect
that any remnants of the Population B that survive in the universe at
the present time would also constitute a special morphological class
of galaxies.

  We propose here that the dwarf spheroidal galaxies may be such
remnants. Dwarf spheroidals are galaxies with luminosities in the
range $10^5 - 10^8 \lsun$, with old stellar populations and devoid of
gas. The velocity dispersion of the stars is often near $10\kms$,
and the surface brightness is usually low. The mass-to-light ratios
are generally much higher than expected for an old stellar population,
even in the center, and they vary over a wide range. Thus, the mass
is dominated by dark matter at all radii, and the inferred central
dark matter densities are higher than in any other known galaxies
(see Gallagher \& Wyse 1994 for a review).
These galaxies have only been studied in the Local Group and a few
nearby clusters, owing to their low luminosity.

  These characteristics are similar to what should be expected in cases
(a) or (b) discussed in Section 2. In this scenario, the dwarf
spheroidals would have been formed in dark matter halos with $\sigma
\sim 10 - 50 \kms$ during reionization. A large fraction of their gas
could have been expelled in a wind, which would have caused the stars to
expand adiabatically in the dark matter potential well as the central
baryonic mass was lost. Most of these galaxies should then have merged
into more massive systems (forming part of the spheroid population of stars
in normal galaxies like the Milky Way; see MR97), but some could
survive until today in orbit in galactic halos, or even have remained
as isolated objects. Many dwarf spheroidals show evidence for multiple
starbursts, some of them having taken place only a few billions years
ago. This does not contradict our picture, but it implies that after
the initial starburst at reionization, dwarf spheroidals were able
to accrete more gas and undergo other starbursts at later times. This
could happen when the accreted gas becomes self-shielding to ionizing
radiation and cools
rapidly, as in the scenario outlined by Babul \& Rees (1992).
Accretion of new gas should occur before these dwarfs are captured in an
orbit within more massive galactic halos, because after this capture
the surrounding gas will be shocked to high temperature and will move
at a high velocity relative to the dwarf spheroidal.

\section{Conclusions}

  Our present understanding of the evolution of the IGM, of its
reionization and its metal enrichment, leads to the expectation that the
first galaxies in the universe were formed at very high redshifts, while
the IGM was not yet fully ionized. These galaxies might have very
different properties from the galaxies we are familiar with in the
present universe; they affected the physical state of the intergalactic
gas that later formed the more massive galaxies like the Milky Way, and
the stars that formed in them are probably part of the spheroidal
components in present-day galaxies. The discovery of these high-redshift
galaxies is a crucial and necessary step to advance to a more complete
understanding of how galaxies formed.

  The requirement of having produced the density of heavy elements
in the \lya forest leads to a prediction of
the mean surface brightness contributed from the Population B galaxies.
This mean surface brightness is not much smaller than that due to the
faintest galaxies so far detected with the HST and ground-based
telescopes. Detection of these first galaxies,
for reasonable estimates of their luminosities,
should be possible with the proposed NGST, which is an ideal tool for
the detection of star-forming galaxies at $ 5 \lesssim z \lesssim 20$
(Mather \& Stockman 1996), using the Gunn-Peterson trough to identify
them. At present, the brightest examples of galaxies at these redshifts
might already be found with HST and Keck, possibly with the help of
lensing magnification in the fields behind rich clusters of galaxies. 

\acknowledgements
We thank Len Cowie, Richard Ellis, Abraham Loeb, Max Pettini,
and Michael Rauch
for useful discussions and for helping us with the numbers for the
sensitivity of the Keck telescope. We thank Chung-Pei Ma, who
suggested to us the nomenclature of Populations A and B, and Bohdan
Paczy\'nski and David Spergel, for discussions on the properties of
dwarf spheroidals. We are grateful to David Weinberg for providing a
code that was used for calculating the halo abundances for Figure 1.

\newpage

\begin{figure}
\centerline{
\hbox{
\epsfxsize=4.4truein
\epsfbox[55 32 525 706]{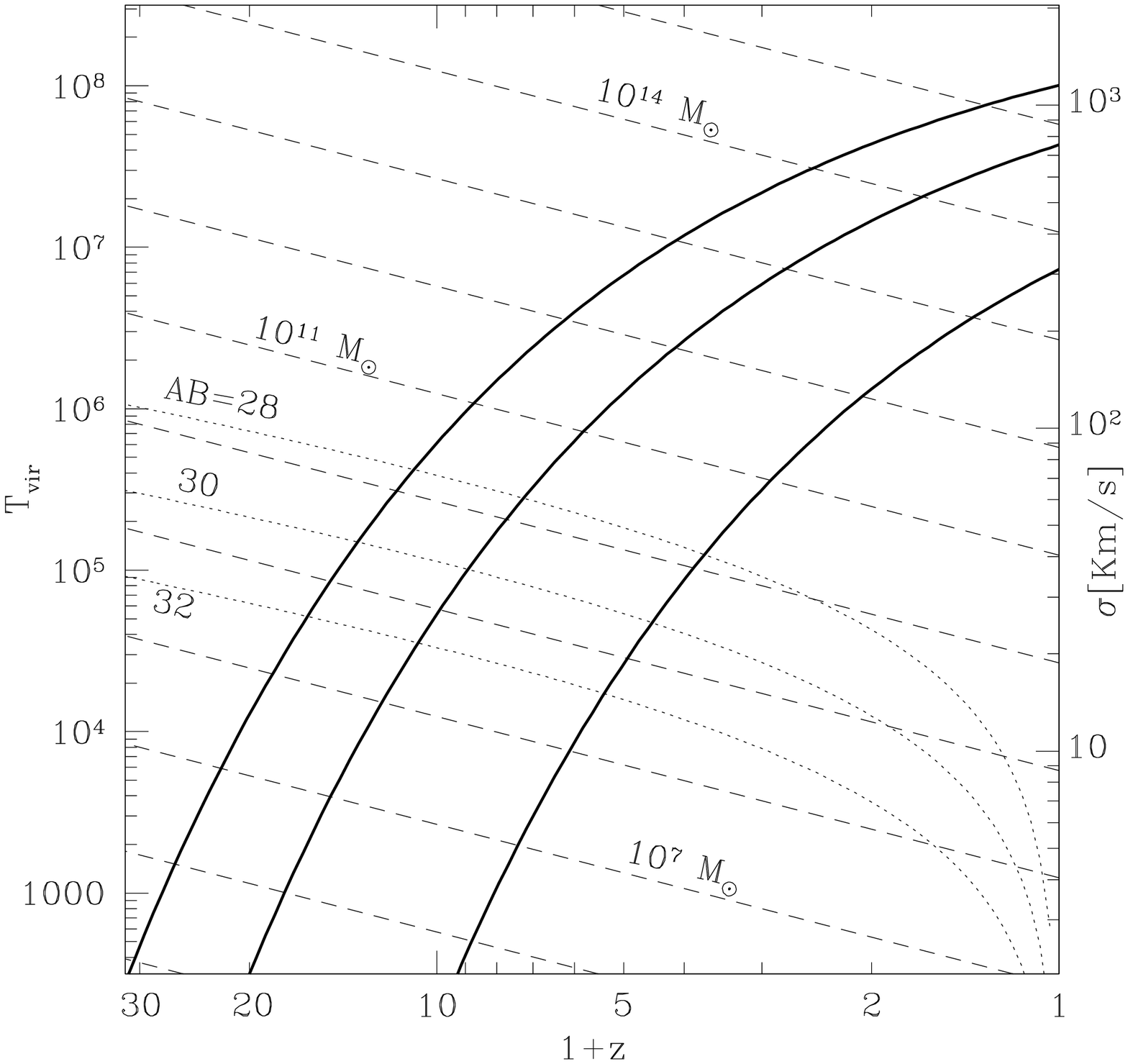}
}
}
\vskip -40pt
\caption{
Thick solid lines give the virial temperature and velocity dispersion
of halos collapsing at redshift $z$, for (1,2,3)-$\sigma$ peaks, in the
cold dark matter model with $\Omega=1$, $h=0.5$, $\sigma_8=0.7$. The
dashed lines are for constant halo mass. The
high-$\sigma$ peaks collapse on a larger scale than the low-$\sigma$
peaks at a fixed epoch, hence they have higher velocity dispersion and
masses. The dashed lines give the AB magnitude of a galaxy formed in
a given halo, assuming that all baryons in the halo (accounting for 8\%
of the total mass) turn to stars in a starburst lasting for $0.3$ times
the age of the universe. The AB magnitude is at a rest-frame wavelength
in the UV, longward of \lya .
}
\end{figure}
\vfill\eject

\end{document}